\documentstyle[preprint,aps,eqsecnum,epsf]{revtex}
\def\beq{\begin{equation}}
\def\eeq{\end{equation}}
\def\ra{\rightarrow}
\def\B{\overline B}
\def\blpkpi{B(\Lambda_c\ra pK^-\pi^+)}
\def\overN{\overline N}
\def\D*{D^{(*)}}
\begin{document}

{\tighten
\preprint{\vbox{
\hbox{FERMILAB--PUB--97/231--T}}}
\title{\bf Heavy Baryon Production and Decay}
\author{Isard Dunietz}
\address{\it Fermi National Accelerator Laboratory, P.O. Box 500,  
Batavia, IL
60510}

\bigskip
\date{\today}
\maketitle
\begin{abstract}
	The branching ratio $\blpkpi$ normalizes the production and decay of charmed and bottom baryons.
At present, this crucial branching ratio is extracted dominantly from $\B\ra$ baryons
analyses. This note questions several of the underlying assumptions and predicts
sizable $\B\ra \D* N\overline N'X$ transitions, which were traditionally neglected.
It predicts $\blpkpi$ to be significantly larger (0.07 $\pm$ 0.02) than the world average. Some consequences are briefly mentioned. Several techniques to measure $\blpkpi$ are outlined with
existing or soon available data samples.
By equating two recent CLEO results, an appendix obtains $B(D^0\ra K^-
\pi^+)= 0.035 \pm 0.002$, which is somewhat smaller than the current world average.
\end{abstract}
\newpage
}
\section{Motivation}

Decays of heavy baryons allow novel tests of Heavy Quark Effective Theory (HQET)~\cite{iw,neubert}. For instance, the structure of the $1/m_c$ corrections is known for the
semileptonic transition $\Lambda_b \to \Lambda_c\ell\nu$~\cite{georgigw}. That
structure is theoretically simpler than the much studied $\overline B\to
D^{(*)}\ell\nu$ one because the light degrees of freedom of the heavy baryon are
spinless and isospinless, while those of the heavy meson are not. Heavy baryon
decays allow one to refine\footnote{The model-independent determination of $|V_{cb}|$ from inclusive semileptonic
$\overline B\to X\ell\nu$ transitions involves corrections dependent on $\overline \Lambda \equiv
m_B -m_b$~\cite{bigiinclusive}. While $\overline \Lambda$ is poorly known at present, it could be determined
more accurately from the $\Lambda_b\to\Lambda_c \ell\nu$ process~\cite{georgigw}.  The theoretical input could thus be  better controlled.} 
the extraction of the fundamental Cabibbo-Kobayashi-Maskawa (CKM) parameters
\cite{georgigw,springer,baryoncp} and could show CP violating effects
\cite{baryoncp}. Detailed studies of heavy baryons are thus important.

The mode $\Lambda_c\to pK^- \pi^+$ plays a central role in those investigations,
because of its sizable branching ratio and observability. Currently, most other
$\Lambda_c$ branching ratios are normalized with respect to $B(\Lambda_c\to
pK^- \pi^+)$. The branching fractions of other weakly decaying charmed baryons
$(\Xi_c,\; \Omega_c)$ can also be tied to $B(\Lambda_c \to pK^-\pi^+)$.
The importance of the $\Lambda_c\to pK^-\pi^+$ process is not limited to the
charm sector but extends to the $b$-sector. Decay products of beautiful
baryons will normally involve charmed baryons.  Even bottom mesons decay
non-negligibly into $\Lambda_c ,\Xi_c ,\Omega_c$ baryons.

The 1996 edition of the particle data group (PDG) quotes~\cite{pdg96}
$B(\Lambda_c \to pK^- \pi^+)=0.044\pm 0.006$, which is dominated by
$\overline B\to$ baryons analyses. The ``traditional" interpretation of the most accurate and recent $\stackrel{(-)}{B} \to \Lambda_c X$ data~\cite{zoeller} leads to a value of $B(\Lambda_c \to pK^- \pi^+)=0.027\pm 0.005$. Those ``traditional" analyses have made simplifying
assumptions which may not hold as discussed below. This note obtains a
significantly larger
$$B(\Lambda_c\to pK^-\pi^+)= 0.07\pm 0.02\;,$$
by combining existing data with theory.

Section 2 discusses the derivation of this sizably larger branching ratio.
Section 3 reviews the traditional extraction of $B(\Lambda_c\to pK^-\pi^+)$
from $\overline B\to$ baryons analyses, and reviews the various
employed assumptions. While the $\overline B\to D^{(*)}N\overline N'X$ transitions were
neglected, a straightforward theoretical Dalitz plot analysis shows that they
probably are sizable~\cite{recal}.  Here $N^{(')}$ denotes a nucleon. This note predicts that
\beq
B(\B\ra D^{(*)}N\overline N'X)\sim \;\;{\rm few}\%\;,
\eeq
and demonstrates that the assumption of neglecting $\B\to\Xi_c X,\Omega_c
X$ cannot be justified.  Finally, the so-called ``model-independent"
determination of \cite{argus92,pdg96}
\beq
B(\B \ra {\rm baryons}) = 0.068 \pm 0.006
\eeq
is questionable as it neglected the $\B\ra D^{(*)}N\overline N'X$ processes. The
latter part of Section 3 suggests several methods to search for and observe
$\B\ra D^{(*)}N\overline N'X$ in existing data samples.
The observation would put into further doubt the conventional $\B\ra$ baryons model. It
would necessitate a serious rethinking of how to accurately determine absolute
branching ratios of heavy baryons. The goal of Section 4 is therefore to sketch several
methods that are able to determine absolute $\Lambda_c$ branching ratios from
existing or soon available data samples. Some implications of the significantly larger predicted
$B(\Lambda_c\ra pK^-\pi^+)$ are discussed in Section 5. Section 6 concludes.

\section{$B(\Lambda_{\lowercase{c}}\ra \lowercase{p}K^-\pi^+)$}

The 1996 particle data group value is~\cite{pdg96}
\beq
B(\Lambda_c\ra pK^-\pi^+) = 0.044\pm 0.006\;.
\eeq
It is dominated by $B\ra$ baryons analyses, which the next section critically
reviews. The recent and more accurate CLEO result \cite{zoeller} would
imply a much reduced
\beq
\label{zoellerbr}
B(\Lambda_c\ra pK^-\pi^+) =0.027\pm 0.005\;,
\eeq
if one were allowed to use the conventional $B\ra$ baryons analysis.

This note argues to use instead
\beq
\label{brtrue}
\blpkpi =0.07\pm 0.02 \;,
\eeq
which is obtained from~\cite{shipsey}
\begin{equation}
\label{shipsey}
B(\Lambda_c\rightarrow pK^-\pi^+)=\frac{B(\Lambda_c\rightarrow
pK^-\pi^+)}{B(\Lambda_c\rightarrow\Lambda X\ell\nu
)}\frac{B(\Lambda_c\rightarrow\Lambda X\ell\nu )}{B(\Lambda_c\rightarrow
X_s\ell\nu )} \frac{\Gamma (\Lambda_c\rightarrow X_s \ell\nu )}{\Gamma (D^0
\rightarrow X_s\ell\nu )}\frac{B(D^0 \rightarrow X\ell\nu )}{\left( 1+\left
|\frac{V_{cd}}{V_{cs}}\right |^2\right)} \frac{\tau (\Lambda_c )}{\tau (D^0 )}.
\end{equation}
The various factors will be discussed in turn.  Experiment informs us about \cite{shipsey}
$$\frac{B(\Lambda_c\rightarrow pK^-\pi^+)}{B(\Lambda_c\rightarrow\Lambda X\ell\nu
)} = 1.93 \pm 0.10 \pm 0.33 \;.$$

Because both the initial state $\Lambda_c$ and the $c \to s \ell \nu$ transition have zero isospin, the resulting final states are isospinless.  Isospin symmetry gives
$$
\Gamma
(\Lambda_c \rightarrow n \overline K^0\ell\nu ) =
\Gamma
(\Lambda_c \rightarrow p K^-\ell\nu ),
$$
$$
\Gamma
(\Lambda_c \rightarrow \Sigma^+ \pi^-\ell\nu ) =
\Gamma
(\Lambda_c \rightarrow \Sigma^- \pi^+\ell\nu ) =
\Gamma
(\Lambda_c \rightarrow \Sigma^0\; [\to \Lambda X]\; \pi^0\ell\nu ) ,
$$
and once applied to the ratio
$f\equiv B(\Lambda_c\rightarrow \Lambda X\ell\nu )/B(\Lambda_c\rightarrow X_s
\ell\nu )$
yields
\begin{equation}
\label{f}
f=1/\left(1+2 \frac{\Gamma (\Lambda_c\rightarrow\Sigma^+ \pi^- \ell\nu )}{\Gamma
(\Lambda_c \rightarrow\Lambda X\ell\nu )} +2\frac{\Gamma (\Lambda_c\rightarrow
pK^- \ell\nu )}{\Gamma (\Lambda_c\rightarrow \Lambda X\ell\nu )} \right)\;.
\end{equation}

The underlying expectation is that the Cabibbo-allowed semileptonic transition $\Lambda_c\ra X_s\ell\nu$ consists almost entirely of $\Lambda X\ell\nu ,\Sigma\pi\ell\nu ,$ and $N\overline
K\ell\nu$. Further, the exclusive $\Lambda_c\ra \Lambda\ell\nu$ transition is predicted to
dominate (almost saturate) $\Lambda_c\ra X_s\ell\nu$, in analogy to what has been observed in $D\ra
X_s\ell\nu$ processes. Cabibbo-allowed semileptonic $D$ decays are basically saturated by
the exclusive $\overline K\ell\nu$ and $\overline K^*\ell\nu$ modes and no evidence for resonant $\overline K^{**}\ell\nu$ or non-resonant $\overline K n\pi\ell\nu (n\geq 1)$ activity has been found~\cite{brd0sl}. Because the
$\overline K^{(*)}$ analogue in the baryon sector is the $\Lambda$ hyperon, the observed $D$
decay pattern indicates a value close to 1 for $f$. Theoretical studies of invariant hadronic mass spectra in $\Lambda_c
\ra X_s\ell\nu$ transitions come to the same conclusion~\cite{dighe}. We thus estimate $f=0.9 \pm 0.1$. Fortunately, $f$ can be determined
experimentally in the future via the right-hand-side of Eq.~(\ref{f}).

The ratio $r\equiv\Gamma (\Lambda_c\rightarrow X_s\ell\nu )/\Gamma
(D^0\rightarrow X_s\ell\nu )$ has been estimated~\cite{bigi}
$$r=1.3 \pm 0.2\;.$$
The prediction for $r$ to be larger than 1 follows from the
operator-product-expansion formalism. The inclusive semileptonic $\Lambda_c$ and
$D^0$ decay rates involve the same leading terms, but differ in the ${\cal
O}
(1/m_c^2)$ corrections~\cite{bigi}. The most significant difference occurs in the average
value of the spin energy. That value vanishes for the $\Lambda_c$, while it
decreases $\Gamma (D\ra X_s \ell\nu )$. That explains why $r$ is expected to be
larger than 1.

CLEO~\cite{brd0sl} gives the most precise $B(D^0\rightarrow X e\nu
)=0.0664\pm 0.0018\pm 0.0029$ measurement to date and the lifetime ratio is
taken from the PDG~\cite{pdg96}.
Eq.~(\ref{shipsey}) expresses the branching ratio of $\Lambda_c\rightarrow pK^-\pi^+$ in terms of Cabibbo
favored transitions, because the $c\rightarrow d$ transitions of $\Lambda_c$
suffer from large Pauli interference enhancements~\cite{voloshin} that are
difficult to quantify. Those enhancements are absent for the semi-leptonic
$D^0$ decays. While phase-space effects for the dominant exclusive transitions
will change the ratio $$\Gamma (D^0\rightarrow X_d \ell\nu )/\Gamma (D^0
\rightarrow X_s \ell\nu )$$
away from the naive estimate $|V_{cd}/V_{cs}|^2$, the expected change will have
negligible effect on the determination for $B(\Lambda_c\rightarrow pK^-\pi^+)$
with present accuracy. Eq.~(\ref{brtrue}) is obtained by combining the above.

\section{Baryon production in $B$ decays}

Because the dominant extractions of $\blpkpi$ \cite{pdg96,zoeller} involve
$\B\ra$ baryons analyses, it is worthwhile to review the various traditional
assumptions made~\cite{cleo92,argus92}. At the present level of accuracy, it is
safe to neglect the $b\ra u$ baryon producing transitions to obtain (see
Fig.~1)
\beq
B(\B\ra \;\;{\rm baryons})= B(\B\ra N_cX) + B(\B\ra D^{(*)}N\overline N' X)\;.
\eeq
Here $N_c$ denotes any weakly decaying charmed baryon $(\Lambda_c ,\Xi_c,
\Omega_c), D$ denotes charmed mesons and $N^{(')}$ stands for a nucleon.
The $\B \ra D^{(*)}N\overline N' X$ processes were traditionally neglected, because of arguments based
on phase space suppression~\cite{cleo92,argus92}. One assumed that
\beq
\label{bar=nc}
B(\B\ra {\rm baryons}) = B(\B \ra N_c X). 
\eeq
Since at the time neither $\Xi_c$ nor $\Omega_c$ production in $\B$ decays were
observed, they were neglected. One thus obtained
\beq
\label{bar=lambdac}
B(\B\ra {\rm baryons})= B(\B \ra \Lambda_c X) .
\eeq

This report distinguishes flavor-specific branching
fractions--$B(\overline B\to TX)$ and $B(\overline B\to \overline
TX)$--from the flavor-blind yield per $B$ decay
\begin{equation}
\label{yT}
   Y_T \equiv B(\overline B\to TX) + B(\overline B\to \overline
TX)\;, 
\end{equation}
where $\overline B$ represents a weighted average of $B^-$ and $\overline B^0$. From flavor-specific and flavor-blind light baryon production in $B$ meson
decays $[\stackrel{(-)}{B} \ra p,p\overline p, \Lambda ,\Lambda \overline p,
\Lambda\overline\Lambda ]$, one deduced that \cite{argus92}
\beq
\label{bbaryons}
B(\overline B\ra {\rm baryons}) = 0.068 \pm 0.006 \;.
\eeq
The $B(\B\ra\Lambda_cX) \blpkpi$ measurement
was then used to obtain $\blpkpi$ by substituting $B(\B\ra \Lambda_c X)$ by the ``measured" $B(\B\ra$
baryons) [Eq.~(\ref{bbaryons})].  The most accurate measurement to date is~\cite{zoeller}
\beq
\label{pkpi}
Y_{\Lambda_c} \times \;\blpkpi = (1.81 \pm 0.22\pm 0.24 ) \times 10^{-3}\;,
\eeq
from which Eq.~~(\ref{zoellerbr}) is obtained.
That summarizes the traditional understanding of baryon production in $B$ meson
decays and the conventional determination of $\blpkpi$.

\subsection{Critique}
In contrast, our
picture of baryon yields in $B$ meson decays is more involved, and $\blpkpi$ is
significantly larger than currently believed. First, are the $\B\ra D^{(*)}N\overline
N'X$ transitions really negligible? A straightforward theoretical Dalitz plot
analysis of the quark subprocess indicates that they probably are sizable. Simple accounting of the
various baryon yields leads independently to the same conclusion. Thus, the
traditional assumption [Eq.~(\ref{bar=nc})] is probably not justified.
Second, is it permissible to neglect $\Xi_c, \Omega_c$ production in $\overline B$ meson
decays [Eq.~(\ref{bar=lambdac})]?  Clearly not, since $\B\ra \Xi_c$ has been
observed \cite{glasgow,browder,gibbons}, and $B(\B\ra \Xi_cX,\Omega_c X)$ has been predicted
\cite{recal} to be a sizable fraction with respect to $B(\B\ra\Lambda_cX)$.

Third, the inclusive $B(\B\ra$ baryons) determination [Eq.~(\ref{bbaryons})]
assumed Eq.~(\ref{bar=nc}). Because that assumption is probably not justified,
the result [Eq.~(\ref{bbaryons})] inferred from flavor-specific light baryon
yields is questionable. Instead of the traditional extraction of $\blpkpi$,
the measurement [Eq.~(\ref{pkpi})] is used to determine the flavor-blind $\Lambda_c $ yield in
$B$ decays,
\beq
\label{ylmc}
Y_{\Lambda_c} =(0.026\pm 0.005) \;\frac{0.07}{\blpkpi}\;.
\eeq
Only about 2.6 \% of all $\B$ decays are seen in modes involving
$\stackrel{(-)}{\Lambda_c}$, in contrast to conventional belief
\cite{argus92,cleo92,pdg96,thorn}.
Before constructing a consistent view of baryon production in $B$ decays, two
apparently puzzling observations are reviewed:

(a) The momentum spectrum of produced $\Lambda_c$ in $\B$ decays is very soft
\cite{zoeller}.

(b) The two-body modes $\B\ra \{\Lambda_c,\Sigma_c \}\; \{\overline
p, \overline\Delta\}$, shown in Figure 2, have not been observed. Only tight upper
limits at the $10^{-3}$ level exist~\cite{pdg96,paulini}.

\subsection{Previous attempt to solve the puzzles}

To resolve these puzzles it was hypothesized that baryon production in $B$
decays is governed by the $b\ra c\overline cs$ transition (see Fig.~3) \cite{dcfw}.
The $\overline \Lambda_c$ momentum spectrum is soft because
the $\overline\Lambda_c$'s are produced in association with the heavy
$\Xi^{(r)}_c$, where superscript ``$r$" denotes resonance.
The two-body modes $\B\ra \{\Lambda_c,\Sigma_c \}\; \{\overline
p, \overline\Delta\}$ are naturally absent.  Further, this mechanism
gives rise to ``wrong-sign" $b\ra\overline\Lambda_c$ transitions in contrast to the
conventional ``right-sign" $b\ra\Lambda_c$ processes. Finally, it followed that
$\Xi_c $ production in $\B$ decays is large and not negligible as commonly
assumed.

Subsequently, CLEO found evidence for a large $\Xi_c$ yield~\cite{browder,glasgow}
\begin{equation}
\label{xic94}
Y_{\Xi_c} = 0.039\pm 0.015\;.
\end{equation}
That same analysis measured the ``wrong-sign" to ``right-sign" $\Lambda_c$
production in $B$ meson decays to be small \cite{glasgow}
\beq
r_{\Lambda_c}\equiv\frac{B(\B\ra\overline\Lambda_cX)}{B(\B\ra\Lambda_cX)}=0.20\pm
0.14\;.
\eeq
This result indicated that the $\B\ra\Xi_c^{(r)}\overline\Lambda_cX$ processes
are not dominant, and refuted the hypothesis that baryon production in $B$ decays is dominated by the $b\ra c\overline cs$ transition.

The flavor-specific $\Xi_c$ and $\Omega_c$
production in $\stackrel{(-)}{B}$ meson decays can be correlated to the much
more accurately measured flavor-specific $\Lambda_c$ yields~\cite{recal}.  For a full list of predictions, please consult Ref.~\cite{recal}.  One prediction is that
\beq
\frac{Y_{\Xi_c}}{Y_{\Lambda_c}}=0.38\pm 0.10 \;,
\eeq
and once combined with (\ref{ylmc}) predicts that
$$Y_{\Xi_c} =(0.010\pm 0.003) \frac{0.07}{\blpkpi}= 0.010\pm 0.004 \;.$$

The much larger
central value quoted by CLEO (\ref{xic94}) indicates that the absolute branching ratio scale
of $\Xi_c$ decays is in truth much larger than assumed. Theoretical
support can be obtained from a recent paper of Voloshin \cite{voloshin}.
Because of the above reasons, the CLEO collaboration now cites~\cite{gibbons}
$$Y_{\Xi_c} = 0.020\pm 0.010 \;.$$
There remains little doubt that $\Xi_c$ production is sizable in $\B$ decays.
Thus the determination of $B(\Lambda_c\ra pK^-\pi^+)$ from previous $\B\ra$
baryons analyses is questionable.

\subsection{Towards a consistent view of baryon production in $B$ decays}

Puzzles (a) and (b) can be explained by noting that a straightforward Dalitz plot
for the dominant $b\ra c\overline ud$ transition predicts the $cd$ invariant mass to be very
large \cite{recal}. (The predicted invariant $cd$ mass distribution follows
from the $V-A$ nature of the $b\ra c\overline ud$ process.) If the $cd$ forms a
charmed baryon [Figure 1], then in general this baryon will be significantly more massive
than a $\Lambda_c$ or $\Sigma_c$, which explains puzzle (b). Further, such very
massive $cdq$ objects or highly excited charmed baryon resonances would be seen usually
as $\Lambda_c n\pi\: (n \geq 1)$. That explains naturally the observed soft $\Lambda_c$
momentum spectrum [puzzle (a)].

Analogously the invariant $cs$ mass in $b \to cs\overline c$ transitions is predicted to be very high.  The $\Xi^{r}_c$ produced in $\overline B \to \Xi^{(r)}_c \overline \Lambda_c X$ processes could be seen significantly as
$\Lambda_c\overline K X\; [\Lambda D X]$ which would lead to $\overline B\rightarrow
\Lambda_c \overline\Lambda_c \overline K X$ $[\Lambda D \overline\Lambda_c X]$ transitions. Such transitions could
comprise a non-negligible fraction of the inclusive $\Lambda_c$ production in
$\overline B$ decays and could show up as $\Lambda_c \overline\Lambda$
correlations in single $\stackrel{(-)}{B}$ decays.\footnote{If $\Lambda_c \overline\Lambda_c$ production in
$B$ decays turns out to be sizable, then the statement of Ref.~\cite{cleo92} that
their $B(\Lambda_c\rightarrow pK^-\pi^+ )$ measurement should be considered
strictly as a lower limit has to be modified.}

While the theoretical Dalitz plot argument predicts the initially produced charmed baryons (via
$b\rightarrow
c)$ to be highly excited, this is not expected of their
pair produced
antibaryons (via $b\rightarrow \overline u$ or $b\rightarrow \overline c)$. The $V-A$
nature of the interaction favors smaller energies for the $\overline u$ or $\overline c$ antiquark in the
restframe of the decaying
$b$. Since the spectator antiquark $\overline q_s$ of the
$\overline B(\equiv
b\; \overline q_s )$ meson involves only a modest Fermi momentum, the invariant mass of
the $\overline u\; \overline q_s$ or $\overline c\; \overline q_s$ system is also expected to be modest.

The very massive $cdq$ produced in $b \to c d \overline u$ transitions could be seen sizably as
$D^{(*)}NX$. The $\B$ meson could be seen therefore in $\B\ra D^{(*)}N\overline N'X$
processes, in contrast to prevailing belief.
Figure 4 shows another $\B\ra D^{(*)}N\overline N'X$ amplitude where the virtual
$W^-\ra\overline ud$ hadronizes into a light baryon antibaryon pair.\footnote{The
size of this amplitude can be estimated from baryon production measurements at
$e^+e^-$ colliders at c.m. energies $\sqrt{s}$ that satisfy $2m_p < \sqrt{s} < m_B-m_D$.}
Sizable $\B\ra D^{(*)}N\overline N'X$ processes would invalidate the assumption that
baryon production involves, in general, weakly-decaying charmed baryons
(\ref{bar=nc}). The current determinations of $\blpkpi$ from $\B\ra$ baryons
analyses would have to be modified.

Another reason why the $\B\ra D^{(*)}N\overline N'X$ processes were neglected is the tight upper
limit \cite{cleo92}
\beq
\label{dstarpp}
B(\B\ra D^{*+} p\overline p X)<0.4\% \;.
\eeq
Our scenario survives, however, because of flavor-correlations \cite{recal}.
Consider the very massive $cdq$ object. It could be seen as a $D^{(*)+}$, which
would normally not be produced in association with a $p$, because 
$$cdq \ra (c\overline d) \;(ddq)=D^{(*)+} \{n,\Delta^{0,-},...\}\;.$$
If a $p$ is required
in the final state, it is more readily correlated with a $D^{(*)0}$ from $cdq$
decays. The virtual $W^-\ra\overline ud$ normally hadronizes as $\overline ud\ra\overline
pn,...$ and may thus survive the constraint of Eq.~(\ref{dstarpp}).
This note predicts that
\beq
\label{DN}
B(\B\ra D^{(*)}N\overline N'X)\sim {\rm few}\%\;,
\eeq
from $\;\; B(\B\ra D^{(*)}N\overline N'X) = B(\B\ra {\rm baryons}) - B(\overline B \ra N_c X).$
First we discuss what can be inferred about inclusive baryon production in $B$
decays, and then we determine $B(\B\ra N_cX)$ \cite{recal}. Prediction (\ref{DN}) follows.

\subsection{$B(\B\ra {\rm baryons})$}

The ``accepted" value~\cite{pdg96,argus92},
\beq
B(\B\ra {\rm baryons})=0.068\pm 0.006\;,
\eeq
is obtained from flavor-specific, light baryon yields,
\beq
\stackrel{(-)}{B}\ra p,p\overline p,\Lambda ,p\overline\Lambda ,\Lambda\overline\Lambda \;,
\eeq
under the assumption that baryon production always involves an $N_c$
[Eq.~(\ref{bar=nc})]. The assumption probably does not hold, raising the
question about the accurate value for $B(\B\ra$ baryons).
Model-independent lower limits can be derived from the light baryon measurements
\footnote{Neutron yields have not been measured yet. The lower limits are obtained
by neglecting them.}
\begin{eqnarray}
\label{barll}
B(\B\ra {\rm baryons}) & \geq & Max \left\{\frac{Y_p}{2},\frac{Y_{{\rm direct}\; p}
+Y_{\Lambda}}{2}, Y_p -B(\B\ra p\overline pX)\right\} = \nonumber \\
& = & Max \left\{ 0.040\pm 0.002, 0.0475\pm 0.0035, 0.055\pm 0.005 \right\}\; .
\end{eqnarray}
Here the flavor-blind yield is defined in Eq.~(\ref{yT}),
and the values summarized in Ref.~\cite{pdg96} were used.

Isospin arguments~\cite{peshkinrosner} could be used to determine $B(\B\ra$ baryons). Consider the dominant
baryon producing transition $b\ra c\overline ud$.
The $\B\ra N$ transition can proceed in several ways, some of which violate isospin
even after the weak decay of the $\B$ (such as the $\B\ra N_c\ra N$ cascades).
Thus, we focus instead on the light antibaryon yield in $\B$ meson decays, $\B\ra \overline
N'$. Because this $\overline N'$ ``contains" the $\overline u$ from the $b\ra c\overline ud$
transition and the other antiquarks forming the $\overline N'$ are as likely
to be a $\overline d$ as a $\overline u$, we expect more $\overline p$ than $\overline n$ production.
Suppose that the ratio of $\overline p/\overline n$ production falls within the range
\beq
\label{pnrange}
1<\frac{B(\B\ra\overline pX)}{B(\B\ra \overline nX)} < 3\;.
\eeq
That range combined with the measurement of \cite{argus92}
\beq
B(\B\ra\overline pX)=0.048\pm 0.004,\;\; {\rm yields}
\eeq
\beq
\label{range}
0.097\pm 0.009 >B(\B\ra {\rm baryons})>0.065\pm 0.006\;.
\eeq
In summary, the inclusive baryon yield is in excess of 0.05 [Eq.~(\ref{barll})],
and probably somewhere in the 0.06 - 0.10 range. Studies of neutron yields in $B$ decays offer one model-independent way to determine $B(\B\ra {\rm baryons})$.

\subsection{$N_c$ production in $\overline B$ decays}

Once the value of $B(\B\ra N_c X)$ is established, the $\B\ra D^{(*)}N\overline N'X$ fraction
can be determined. The $B(\B\ra N_c X)$ is determined in two steps,
\beq
B(\B\ra N_cX)= B(\B\ra\Lambda_cX) +\; [(B(\B\ra\Xi_cX)+B(\B\ra\Omega_cX)]\; .
\eeq
The flavor-specific $B(\B\ra\Lambda_cX)$ is taken from experiment~\cite{zoeller,glasgow}, whereas
$B(\B\ra\Xi_c X)$ and $B(\B\ra\Omega_c X)$ are correlated\footnote{This note does not use the direct $\Xi_c$ measurement in $\B$ decays, because of
the very large uncertainties involved (see above). Instead, the uncertainties are
drastically reduced by correlating the predicted $\Xi_c$ and $\Omega_c$ yields to
the better measured $\Lambda_c$ yield.} to the observed
$\Lambda_c$ yields~\cite{recal} and therefore are predictions dependent on $B(\Lambda_c\ra pK^-\pi^+)$,
\beq
\label{bnc}
B(\B\ra N_cX)=(0.032\pm 0.006) \frac{0.07}{\blpkpi}= 0.032\pm 0.011 \;.
\eeq
The last equation follows by inserting 0.07 $\pm$ 0.02 for $\blpkpi$. Since the
inclusive baryon yield is at the 0.06-0.1 level and $B(\B\ra N_cX)$ is given by
(\ref{bnc}), we predict that probably
\beq
B(\B\ra D^{(*)}N\overline N'X)\sim {\rm few}\%.
\eeq
Our prediction gets additional support from the theoretical Dalitz plot argument outlined above~\cite{recal}.
Table 1 summarizes this section. The existing data samples are
sufficiently large to search for and observe sizable $\B\ra D^{(*)} N\overline
N'X$ processes.

\subsection{$\B\ra D^{(*)}N\overline N'X$ Search}

Here we list a few suggestions to search for $\B\ra D^{(*)}N\overline N'X$. At the
$\Upsilon (4S)$ one could look for non-trivial angular correlations between a
$D^{(*)}$ and a light baryon or antibaryon $\stackrel{(-)}{N}$. If the $D^{(*)}$
comes from one $\B$ and the $\stackrel{(-)}{N}$ from the other $B$, then they
should be distributed almost isotropically~\cite{tipton}. On the other hand,
observing a nontrivial angular correlation would indicate single $\B$ parentage.
It is likely, however, that either the $D^{(*)}$ or the $\stackrel{(-)}{N}$ or
both are soft, in which case the angular correlations are largely lost.
Thus we recommend to search for the triple correlation $\ell^+
D^{(*)}\stackrel{(-)}{N}$ on the $\Upsilon(4S)$~\cite{recal}. The $\ell^+$ and the $D^{(*)}$
cannot originate from a single $\stackrel{(-)}{B}$ meson, because of their flavors. Either the
$\stackrel{(-)}{N}$ and the lepton share $B$ parentage or the $\stackrel{(-)}{N}$ and the $\D*$
originate from the same $\B$. While the latter interpretation is our coveted
signal, 
the former is very unlikely.  The former interpretation would indicate semileptonic
$B$ decay in conjunction with baryon/antibaryon production\footnote{A sufficiently large lepton 
momentum removes any background
from $\B\ra\Xi_c\overline\Lambda_c X$, where one of the charmed baryons decays
semileptonically and either one contributes the $\stackrel{(-)}{N}$.}
\beq
B\ra\overN_c N'\ell^+\nu \;,
\eeq
which is expected theoretically to be tiny, and for which tight upper limits
already exist \cite{pdg96,lambdacl}.

Bottom hadrons produced at $Z^0$ factories are boosted and hadronize generally in
opposing hemispheres. After selecting a $b$-enriched event, one could search, in a
single $b$-hemisphere, for the predicted few percent (detached) $\D*\overN$
correlation.\footnote{The small ($\sim 0.5\%$) background from
$\overline\Lambda_b\ra\overN_c [\ra\overN X]\D* KX\Longrightarrow \overN\D*$ processes can be
significantly reduced by (a) flavor-tagging, which enriches $b$ over $\overline b$
content, (b) enhancing $B^-$ parentage via vertex charge, which reduces the
$b$-baryon background thereby making even the (detached) $\D* N$ correlations a convincing
signal.}

At either $e^+e^-$ or hadron colliders, one  may attempt to  reconstruct the
$\B\ra \D* N\overN ' \pi$'s modes. The existence of a $\stackrel{(-)}{n}$ can be
inferred in analogy to methods developed for $\nu$ or $K_L$ reconstruction.
Sizable $\B\ra \D* N\overN 'X$ processes would further question the traditional
$\blpkpi$ determinations. The next section outlines briefly some methods that
allow the determination of $\blpkpi$ from existing or soon available data samples.

\section{On determining $B(\Lambda_{\lowercase{c}}\ra \lowercase{f})$}

This section lists a few methods that allow the determination of $B(\Lambda_c\ra
f)$, where $f$ denotes an exclusive $(pK^-\pi^+,\Lambda\pi ,...)$ or
semi-inclusive $(\Lambda X, pX,...)$ $\Lambda_c$-mode.

\subsection{Method (a):}
At  $e^+e^-$ or $p\overline p$ colliders, produce $\Lambda_c\overline\Lambda_c$ pairs at
threshold. Fully reconstruct one of the charmed baryons. Then one determines
$B(\Lambda_c\ra f)$, by measuring the probability for the remaining $\Lambda_c$ to be seen in $f$~\cite{alam80}.

\subsection{Method (b):}
At fixed target experiments, the production asymmetry can be used to determine
$B(\Lambda_c\ra f)$~\cite{cheung}. Since the total produced number of charm quarks
equals that of anticharm quarks, one obtains
\begin{eqnarray}
N(\Lambda_c) -N(\overline\Lambda_c) & + & N(\Xi_c,\Omega_c) -N(\overline \Xi_c, \overline \Omega_c) =
\nonumber \\
& = & N(\overline D) -N(D) +N(\overline D_s) - N(D_s),
\end{eqnarray}
where $N$ denotes the total produced number.
In the lack of a $\Xi_c,\Omega_c$ production asymmetry, the coveted absolute
$B(\Lambda_c\ra f)$ is obtained via
\beq
B(\Lambda_c\ra f)=\frac{N(f) -N(\overline f)}{N(\overline D )-N(D)+N(\overline D_s)
-N(D_s)}\;.
\eeq
If a $\Xi_c, \Omega_c$ production asymmetry is observed, it can be incorporated to
determine $B(\Lambda_c\ra f)$.

\subsection{Method (c):}
The probability that a leading s-quark jet hadronizes as a hyperon $[P(s\ra$
hyperon)] can be experimentally measured. The relevant (diquark) parameters in
current simulation models could then be tuned to agree with the measurements. The
simulation model then predicts the probability for charmed baryon production
$[P(c\ra N_c)]$, with uncertainties typically at the (10 - 20)\% level. Since the charm production
cross-section is known, reconstructing final states of $\Lambda_c\;\;(f)$ permits the
extraction of $B(\Lambda_c\ra f)$.

\subsection{Method (d):}
 At an $\Upsilon (4S)$ factory, reconstruct a $\overline c$-hadron
$[D^{(*)-},\overline D^0,\overline\Lambda_c ]$ with sufficiently high momentum from the
continuum to remove the $B\overline B\ra\overline cX$ background.\footnote{No momentum cut is required when the data is taken at the continuum
below the $\Upsilon (4S)$ resonance.}
The ``opposite" hemisphere contains in general a $c$-hadron. Determine the number
of events that an antiproton is made in the ``opposite" hemisphere $N[\overline c\; \overline
p]$. This $\overline p$ cannot be  a decay product of the $c$-hadron and indicates
$c$-baryon production in the ``opposite" hemisphere.  The observation of
$\Lambda_c$-modes $f$ in the ``opposite" hemisphere allows the measurement of
$N[\overline c\; \overline p f]$ and of
\beq
\label{br4s}
B(\Lambda_c\ra f)\approx\frac{N[\overline c\; \overline pf]}{N[\overline c\; \overline p]}\;.
\eeq
A few corrections and comments must be made before this method becomes promising.

While the existence of an ``opposite" hemisphere $\overline p$ indicates $c$-baryon
production, one must correct for the fraction of the time the $\overline c\; \overline
p$-correlation occurs with $c$-meson production.  That correction can be
determined by measuring the $c$-meson yield in the ``opposite" hemisphere, $N[\overline
c\; \overline p D_{(s)}].$
The probability that $\overline c\; \overline p$ opposite hemisphere events contain $c$-baryons
can thus be determined. While the dominant fraction is $\Lambda_c$'s, one may wish to
correct for the much smaller $\Xi_c ,\Omega_c$ yields.\footnote{The $\Xi_c$ and
$\Omega_c$ yields can be measured. Rates of
specific $\Xi_c$ and $\Omega_c$ modes are related to
specific $\Lambda_c$-modes by the SU(3)-flavor symmetry~\cite{dattado}. The rates of those specific modes are
normally measured well with respect to the calibrating modes. Thus, the fraction
of $\Lambda_c, \Xi_c, \Omega_c$ in $\overline c\; \overline p$ events can be determined.}

    It appears that this method may not yet be feasible, because of the poor
statistics for the various triple correlations $[\overline c\; \overline p f,\overline c\; \overline
pD_{(s)}, \overline c\; \overline p\{\Xi_c,\Omega_c\}]$.
It is thus important to note that much larger statistics are involved in ``single" hemisphere $\overline p f,\overline p D_{(s)}, \overline p\{\Xi_c, \Omega_c\}$ events. The
existence of a $\overline c$-hadron (normally) in the other hemisphere can be
inferred from the reconstructed $c$-hadron~\cite{lewis}. Thus, the poor statistics
of the triple correlations can be avoided, and $B(\Lambda_c\ra f)$ can be measured~\cite{lewis}.\footnote{This $B(\Lambda_c\ra f)$ measurement neglects long range
$c\overline c$ production correlations, which are expected to be small. They can be
accounted for in triple correlation studies.}

\subsection{Method (e):}
This method requires a superb vertex detector. After selecting a $b$-sample, the
sample of fragmentation $\overline p$'s, which originate from the interaction point and
are close to the $b$-jet, indicate $b$-baryon production. (Below we discuss how
to correct for $b$-meson production in association with such $\overline p$'s.)
Observe also negative leptons $\ell^-$ with high $p_T$ and significant impact
parameter, which normally are primary decay products of $b$-decays. The produced
number of such $\overline p\ell^-$ correlations $N[\overline p\ell^-]$ is proportional to
\beq
\label{pbarl}
N[\overline p\ell^-]\sim P(...\ra\overline p)
P(b\ra\Lambda_b)B(\Lambda_b\ra\Lambda_cX\ell\nu )\;,
\eeq
where we assumed that semileptonic $\Lambda_b$ decays are almost always
accompanied by a $\Lambda_c$, apart from tiny
$\Lambda_b\ra \{\Xi_c KX, D^{(*)}NX,...\}\ell\nu$ and  $b\ra u$ processes. The tiny processes are at most at the 10\% level of inclusive semileptonic $\Lambda_b$ decays, as can be inferred from an analogy to semileptonic $\overline B-$decay measurements~\cite{warsawaleph,isgur}.
Ratios of specific $\Lambda_c$ decay rates can thus be determined 
\beq
\label{ratioffp}
\frac{\Gamma (\Lambda_c\ra f')}{\Gamma (\Lambda_c\ra f)} = \frac{N[\overline p\ell^-
f']}{N[\overline p\ell^- f]}\;,
\eeq
where $f, f'=pK\pi ,\Lambda n\pi ,\Lambda X,pX,...$.
Information concerning semi-inclusive $\Lambda_c$ decay rates can thus be obtained.
Even absolute $\Lambda_c$ branching ratios can be determined via
\beq
\label{brpbarl}
B(\Lambda_c\ra f)\approx \frac{N[\overline p\ell^- f)}{N[\overline p\ell^-]}\;.
\eeq
One must correct for the fraction of the time the $b$-jet
nearby the fragmentation $\overline p$ gives rise to a $b$-meson (rather than a
$b$-baryon). The correction factor can be obtained in several ways:
\begin{enumerate}
\item Measure the probability that $\overline p \ell^-$ events involve (detached)
charmed mesons, which together with the $\ell^-$ point to the $b$-decay vertex.
\item Select a charged $B^-$ sample by using vertex charge and vertex mass.
Study the fraction of the time this $B^-$ sample involves a nearby fragmentation
$\stackrel{(-)}{p}$.
\item In the $b$-enriched sample with one fragmentation $\overline p$, observe an
additional fragmentation $p$ also from the interaction point and nearby the
$b$-jet. Since baryon number is conserved, it is likely a
$b$-meson was produced. (The possibility of more than one nearby baryon-antibaryon
pair production may not be negligible, however.) Because isospin symmetry allows one
to determine the production ratio of fragmentation neutrons versus protons, the
fraction of the time $b$-mesons are made can be inferred.
\end{enumerate}
It may prove useful to introduce stringent cuts on the fragmentation $\overline p$,
so as to reduce the $b$-meson fraction.

\subsection{Method (f):}
If it were possible to theoretically relate $\Gamma
(\Lambda_b \ra\Lambda J/\psi )$ to $\Gamma (\B\ra \overline K^{(*)} J/\psi)$
then the observed $\Lambda J/\psi$ sample of fully reconstructed $\Lambda_b$
decays permits the determination of $P(b\ra\Lambda_b )$.
Once $P(b\ra\Lambda_b )$ is known, then the $\Lambda_c [\ra f]\ell^-$ sample
yields $B(\Lambda_c \ra f)$.

\subsection{Conclusion}
Those are then some suggestions to determine $B(\Lambda_c \ra f)$.
Undoubtedly, many possible variations and improvements will become obvious to the
dedicated experimenter. Appendix A sketches a determination of $B(\Lambda_b\ra
X\ell\nu )$ and $|V_{cb}|$ with a reduced dependence on $B(\Lambda_c \ra f)$.

\section{Implications}

If $\blpkpi$  turns out to be significantly larger than the current
world average, as we predict, then there will be many
ramifications. Some of them are:
\begin{itemize}
\item Since $\blpkpi$ normalizes most heavy baryon productions and decays, the
heavy baryon decay tables listed in Ref.~\cite{pdg96} will have to be recalibrated
accordingly.
\item The $\Lambda_b :\B :\B_s$ production fractions will be affected.
The $\Lambda_b$ fraction will be reduced sizably from current estimates while the
$\B$ and $\B_s$ fractions will increase. The measured $\Lambda_b$ branching ratios thus
increase sizably, while the $B_{(s)}$ ones decrease.
\item The number of charms per $b$-decay decreases on two counts. First, the
$\Xi_c$ yield in $\B$ decays is predicted to be sizably lower than its measured central value. Second,
the $\blpkpi$ is significantly larger than expected, resulting into a lower
$\Lambda_c$ yield in $b$ decays than presently believed. Quantitative estimates can be found in Refs.~\cite{recal,disy}.
\item The charmless yield in $B$-meson decays is larger than a recent indirect
extraction \cite{thorn}. CLEO \cite{thorn} measured the flavor-specific
charm yields in $B$ decays in a way that removes the large systematic  uncertainty
due to $B(D^0\ra K^- \pi^+ )$.\footnote{That study enables Appendix B to point out a complementary method for determining $B(D^0\ra K^- \pi^+ )$, which does not involve the soft pion from $D^{*+} \to \pi^+ D^0$ transitions.  The result is $B(D^0\ra K^- \pi^+ ) =  0.035 \pm 0.002$, by equating two CLEO measurements for $Y_D$.}
That beautiful analysis then assumed that the inclusive baryon yield in $\B$ decays is saturated by
weakly decaying charmed baryon production $(N_c)$, resulting in $B(b\ra$ no open $c) = 0.04
\pm 0.04$. In contrast, this note questions the validity of the assumption and uses $B(\B\ra N_c X) = 0.032 \pm 0.011$, which implies a
larger $B(b\ra$ no open $c) = 0.07 \pm 0.04$.  Table II summarizes both viewpoints.
\end{itemize}

Those are then some of the consequences of our view of heavy baryon production
and decay.

\section{Conclusions}

Because the $\Lambda_c\ra pK^-\pi^+$ process normalizes heavy baryon production and
decay, its absolute branching ratio must be known to the highest accuracy
achievable. Analyses of baryon production in $B$ meson decays dominate the
traditional $\blpkpi$ ``measurements",
\begin{eqnarray}
\blpkpi = \left\{ \begin{array}{cc}
 0.044 \pm 0.006 & \cite{pdg96}, \\
0.027 \pm 0.005 & \cite{zoeller}.
\end{array}
\right.
\end{eqnarray}
Those analyses however made several questionable assumptions, summarized in Table
1. Instead, a considerably larger $\blpkpi$ emerges
\cite{shipsey,recal,hawaii}, 
$$
\blpkpi = 0.07 \pm 0.02,
$$
by combining theory and available experimental data on semileptonic charm transitions.

This confusing state of affairs can be clarified by searching for and observing
the traditionally overlooked $\B\ra\D* N\overline N'X$ processes.
We predict those $\B\ra\D* N\overline N'X$ processes to constitute a sizable
fraction of all $\B\ra$ baryons transitions. They should be observable in existing data samples.
If this is borne out, then $\blpkpi$ must be determined afresh. That can be
accomplished in a variety of methods, some of which Section 4 briefly outlined.  A
considerably larger $\blpkpi$ than currently accepted will have ramifications, such
as:
\begin{itemize}
\item The heavy baryon decay tables will have to be recalibrated.
\item The $b\ra\Lambda_b$ production fraction decreases.
\item The measured number of charm per $b$-decay
decreases.
\item The $\B\ra$ baryons transitions are more involved than currently modeled.
\end{itemize}
More theoretical and experimental investigations are highly welcome, as it will improve our understanding of heavy baryon production and decay.

\acknowledgements

We thank M. Barnett for encouraging us to write up this note.  We are grateful to M.S.~Alam, M.~Beneke, D.Z.~Besson, G.~Buchalla, P.~Burchat, R.N.~Cahn, H.W.K.~Cheung, Su Dong, O.~Hayes, N.~Isgur, M.~Luke, M.~Paulini, J.L.~Rosner, I.~Shipsey, N.G.~Uraltsev, M.~Voloshin and C.~Wohl for discussions. This work was supported in part by the Department of Energy, Contract No.
DE-AC02-76CH03000.

\appendix
\section{On Determining $B(\Lambda_{\lowercase{b}}\ra X\ell\nu )$ and $|V_{\lowercase{cb}}|$ with a reduced
$B(\Lambda_{\lowercase{c}}\ra \lowercase{f})$ uncertainty.}

At present there exists a discrepancy between theory and experiment concerning the
lifetime ratio $\tau (\Lambda_b )/\tau (B_d)$ \cite{pdg96}.
The most plausible explanation is an enhanced nonleptonic $\Lambda_b$-rate,
without a corresponding enhancement in the semileptonic $\Lambda_b$-rate.
This Appendix discusses one way to determine the semileptonic $\Lambda_b$ rate and branching ratio,
thereby probing the underlying cause of the lifetime discrepancy.
Building on the discussion presented in method (e) of Section 4, one notes that
the produced number of $\ell^- f$ events is proportional to
\beq
N[\ell^- f]\sim P(...\ra b) P(b\ra\Lambda_b )B(\Lambda_b \ra\Lambda_c
X\ell\nu ) B(\Lambda_c \ra f) \;,
\eeq
where it is understood that the various backgrounds have been corrected for.
In addition, the direct production of $\Lambda_c$ baryons can be studied as well.
To remove the large $b\ra\Lambda_c $ background, it may prove advantageous to
focus on the high momentum $\Lambda_c\ra f$ sample,
\beq
N[f]\sim P(...\ra c) P(c\ra\Lambda_c ) B(\Lambda_c\ra f) \;.
\eeq
The production ratio $P(...\ra b)/P(...\ra c)$ is well known, and HQET can in
principle determine the ratio\footnote{It would be useful to calculate the momentum
dependence of this ratio.}
\beq
\frac{P(b\ra\Lambda_b )}{P(c\ra\Lambda_c )} = 1+ {\cal O} (1/m_Q )\;.
\eeq
The semileptonic $\Lambda_b$-branching ratio can thus be determined
\beq
B(\Lambda_b \ra X\ell\nu )\approx B(\Lambda_b \ra \Lambda_c X\ell\nu )\sim
\frac{N[\ell^- f]}{N[f]} \;.
\eeq

Even the CKM parameter $|V_{cb}|$ can be extracted by measuring the
\underline{exclusive} $\Lambda_b\ra\Lambda_c \ell\nu$ branching ratio,
$B(\Lambda_b\ra \Lambda_c\ell\nu )$. That measurement combined with $\tau
(\Lambda_b )$ determines $\Gamma (\Lambda_b\ra\Lambda_c \ell\nu )$. HQET
\cite{georgigw,neubert} and lattice studies~\cite{kenway} inform on the relevant form-factors, so
that $|V_{cb}|$ can be determined from that $\Lambda_b \ra\Lambda_c \ell\nu$
measurement.

\section{On the  $B(D^0\ra K^-\pi^+)$ Value}

Not only does $B(D^0\ra K^-\pi^+)$ calibrate most other $D^0$ decays, but the $D^+$ and $D_s^+$ calibration-modes [$D^+ \to K^- \pi^+ \pi^+$ and $D_s^+ \to \phi \pi^+$] are tied to it as well~\cite{pdg96}.  The $D^0\ra K^-\pi^+$ process normalizes charmed meson production and decay, in analogy to the role of the $\Lambda_c \to p K^- \pi^+$ transition for charmed baryon studies.    This Appendix introduces one more method for determining $B(D^0\ra K^-\pi^+)$, which yields
\beq
\label{d0kpi}
B(D^0\ra K^-\pi^+)=0.035 \pm 0.002.
\eeq

There exist now two CLEO measurements of $D$ production in flavor-blind $B$ decays,
\beq
Y_D\equiv B(\B\ra DX)+B(\B\ra\overline DX)\;.
\eeq
The first is a high statistics measurement of inclusive $D$ and $\overline D$ production in $B$ decays, which is
inversely proportional to $B(D^0\ra K^-\pi^+)$ \cite{gibbons},
\beq
Y_D = (0.876 \pm 0.037) \left[\frac{0.0388}{B(D^0\ra K^-\pi^+)}\right]\;.
\eeq
The second is not sensitive to $B(D^0\ra K^-\pi^+)$,
\beq
\label{yDthorn}
Y_D = 0.96 \pm 0.05\;,
\eeq
and was
deduced from Ref.~\cite{thorn} as discussed below. Equating the two and
solving for $B(D^0\ra K^-\pi^+)$ yields (\ref{d0kpi}).
This value agrees with the one obtained in
Refs.~\cite{recal,distw,montreal}, and is somewhat below the world average~\cite{richman}, 
$B(D^0\ra
K^-\pi^+) = 0.0388\pm 0.0010$.

The world average is dominated by studies involving the soft $\pi^+$ in $D^{*+}
\ra\pi^+D^0$ decays, which require the accurate modeling of the tails of the soft pion momentum spectrum. Because such accurate modeling may prove more difficult than presently appreciated~\cite{montreal}, measurements of $B(D^0\ra K^-\pi^+)$ insensitive to such soft $\pi^+$'s should also be pursued. Such methods were discussed in the literature~\cite{recal,montreal,thorn}.  This Appendix introduces yet another one.

Eq.~ (\ref{yDthorn}) is obtained via
\beq
Y_D = D_\ell \times (1 + r_D) \times L.
\eeq
Here theory delivers~\cite{thorn}
$$L \equiv B(\B \to D X \ell \nu) / B(\B \to X \ell \nu) = 0.97 \pm 0.02,$$
while the other quantities were measured~\cite{thorn}:
$$D_\ell \equiv \frac{B(\B \to D X )} {B(\B \to D X \ell \nu) / B(\B \to X \ell \nu)} = 0.901 \pm 0.037\;,$$
$$r_D \equiv B(\B\ra \overline DX)/B(\B\ra DX) = 0.100 \pm 0.031\;.$$
Note that several uncertainties cancel in the $D_\ell \times (1 + r_D)$ combination~\cite{thorndikeprivate}, so that CLEO can determine $Y_D\;[B(D^0\ra K^-\pi^+)]$ with a smaller error than given in Eq.~(\ref{yDthorn}) [(\ref{d0kpi})].

\begin{table}
\caption{The conventional determination of $\blpkpi$ involves $\B\ra$ baryons
analyses with the assumptions tabulated here.}
\begin{tabular}{|l|l|}
Assumptions & Comments \\
\hline
\hline
$B(\B\ra D^{(*)}N\overline N'X)=0 $ & Could be $\sim$ few \%. Theoretical Dalitz plot allows
for  \\
& sizable $\B\ra D^{(*)}N\overline N'X$.
\\ 
\hline
$B(\B\ra\Xi_c X,\Omega_cX)=0$ &  $\B\ra \Xi_cX$ large \\
\hline
$B(\B\ra {\rm baryons}) =0.068\pm 0.006$ & Value questionable, because derived under the
assumption \\
& that $B(\B\ra D^{(*)}N\overline N'X)=0.$ Instead, this note \\
&  determines \underline{model-independent lower limits} \\
& from existing light baryon yields.  \\ 
\end{tabular} 
\end{table}

\begin{table}
\caption{Open charm production in $\overline B$ decays. The last row lists the charmless yield which is obtained via
$B(b\ra $no open c)$=1-B(b\ra$ open $c)$.}
\begin{tabular}{|l|c|}
$\overline B[\equiv b\overline q] \ra$ open $c$  & CLEO \cite{thorn} (This note) \\ 
\hline
$\overline B\ra D^0,D^+$ & 0.87 $\pm$ 0.04 \\
\hline
$\overline B\ra D^+_s$ & 0.02 $\pm$ 0.01 \\
\hline
$\overline B\ra N_c\;[\equiv\Lambda_c ,\Xi_c ,\Omega_c ]$ & 0.065 $\pm$ 0.015 
(0.032 $\pm$ 0.011) \\
\hline
$\overline B\ra$ no open $c$ & 0.04 $\pm$ 0.04 (0.07 $\pm$ 0.04) \\
\end{tabular} 
\end{table}

\begin{figure}
\epsfysize = 3in
\centerline{\vbox{\epsfbox{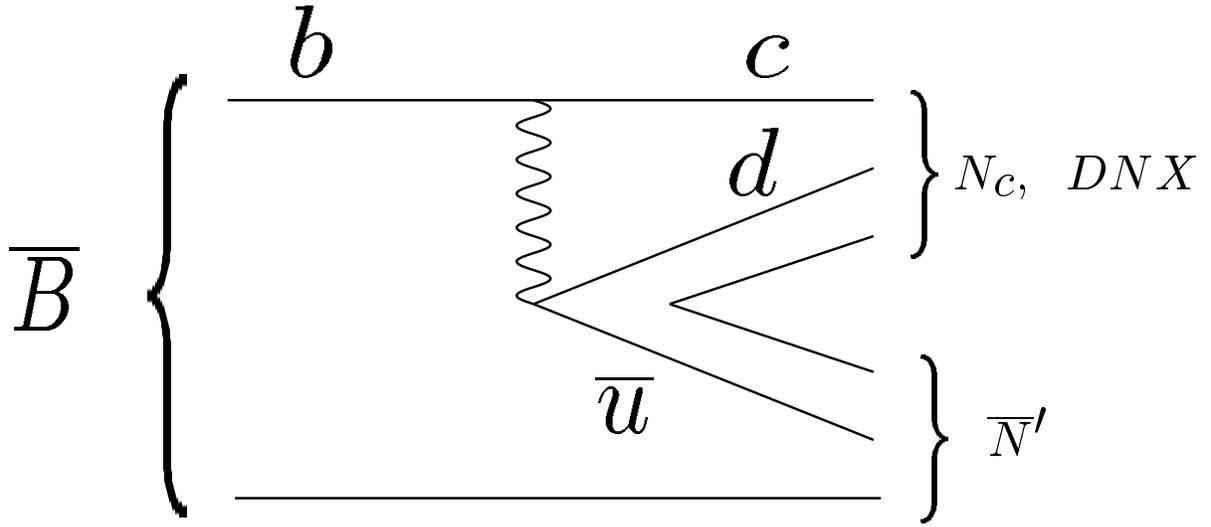}}}
\caption{Graph depicting baryon production in $\B$ decays.}
\end{figure}

\begin{figure}
\epsfysize = 3in
\centerline{\vbox{\epsfbox{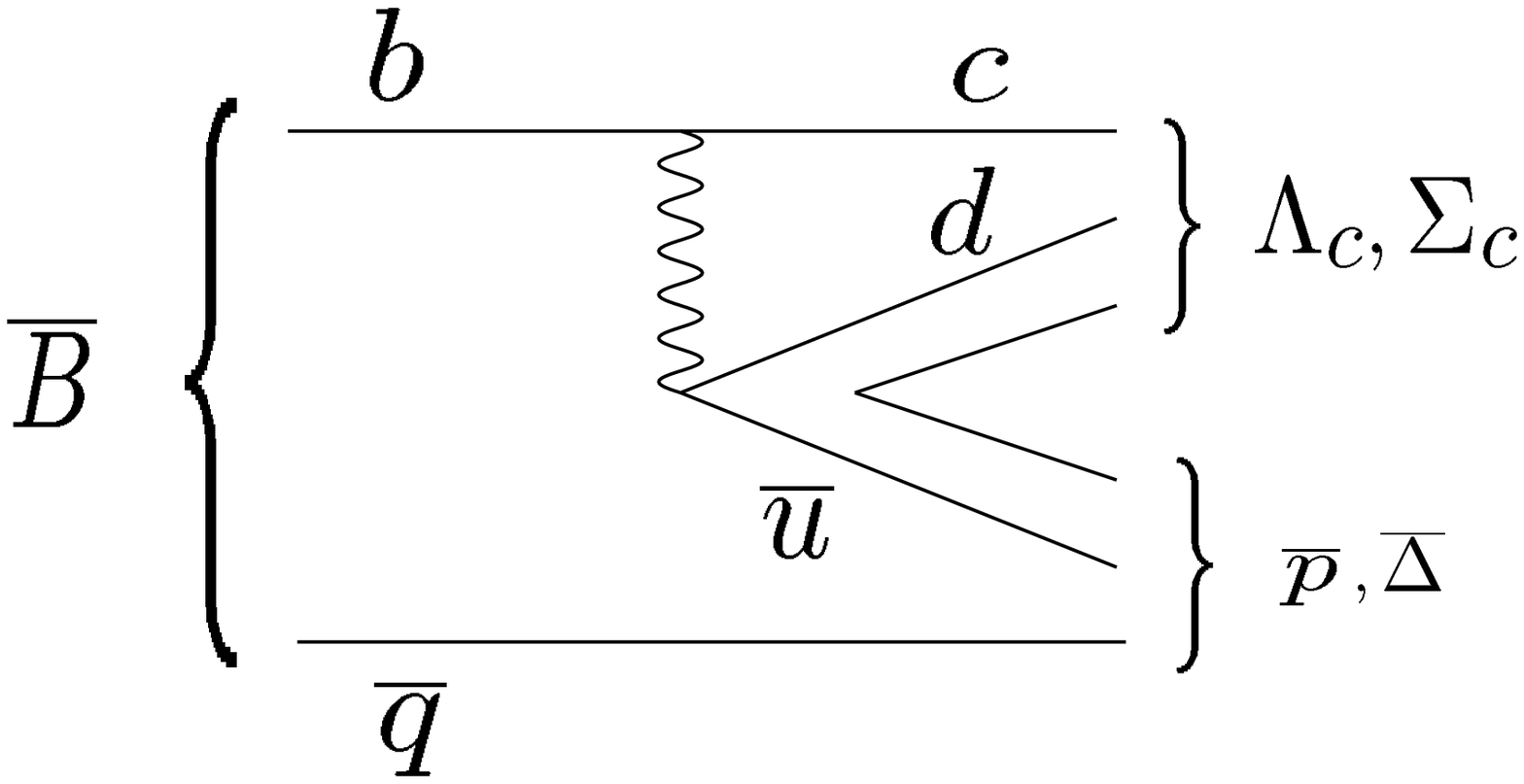}}}
\caption{Graph responsible for the two-body $\B$-modes, $\B\ra
\{\Lambda_c,\Sigma_c\}\{\overline p, \overline\Delta \}$.}
\end{figure}

\begin{figure}
\epsfysize = 3in
\centerline{\vbox{\epsfbox{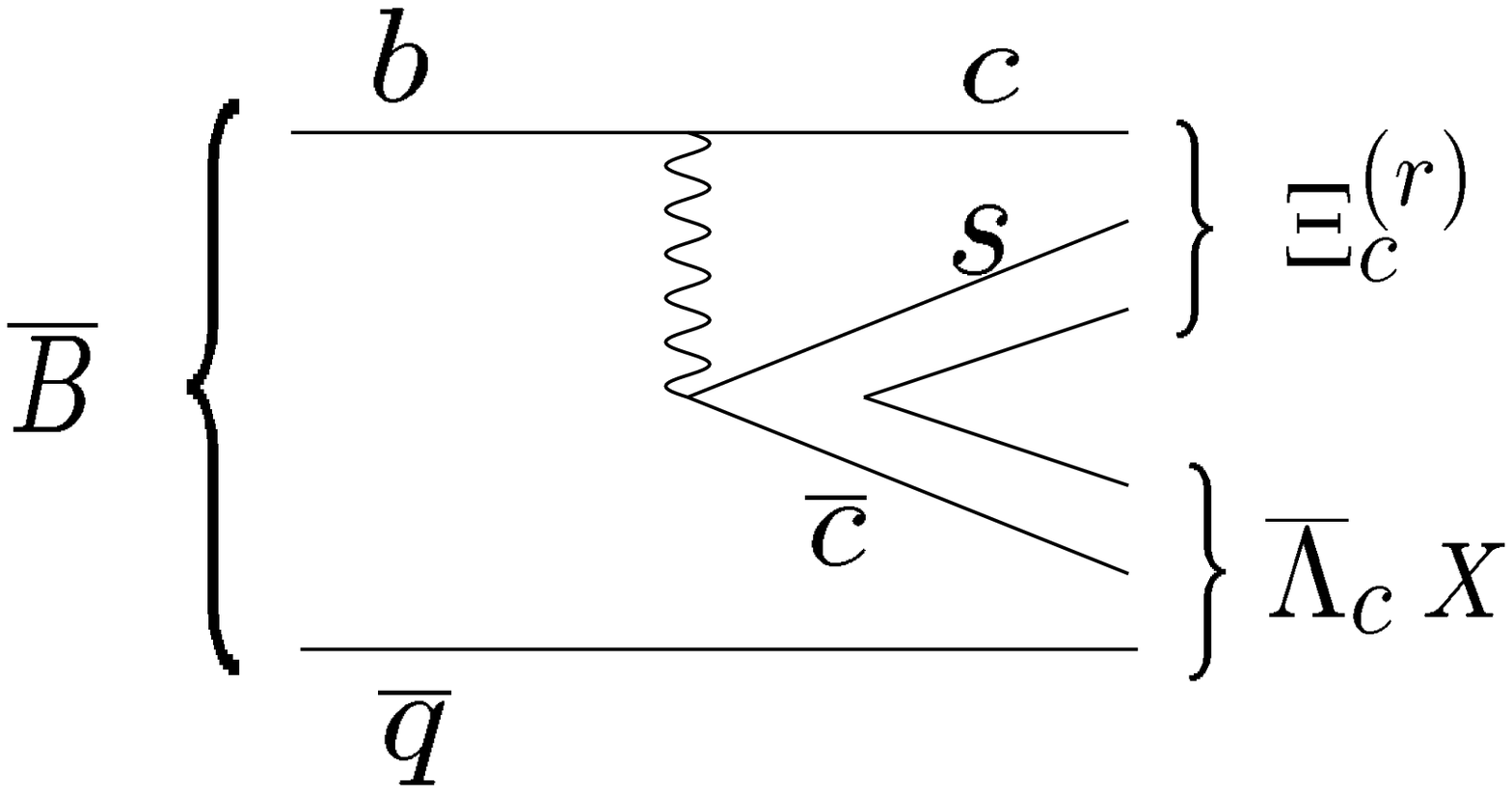}}}
\caption{Graph governing $\B\ra\Xi_c^{(r)}\overline\Lambda_c X$ transitions.}
\end{figure}

\begin{figure}
\epsfysize = 3in
\centerline{\vbox{\epsfbox{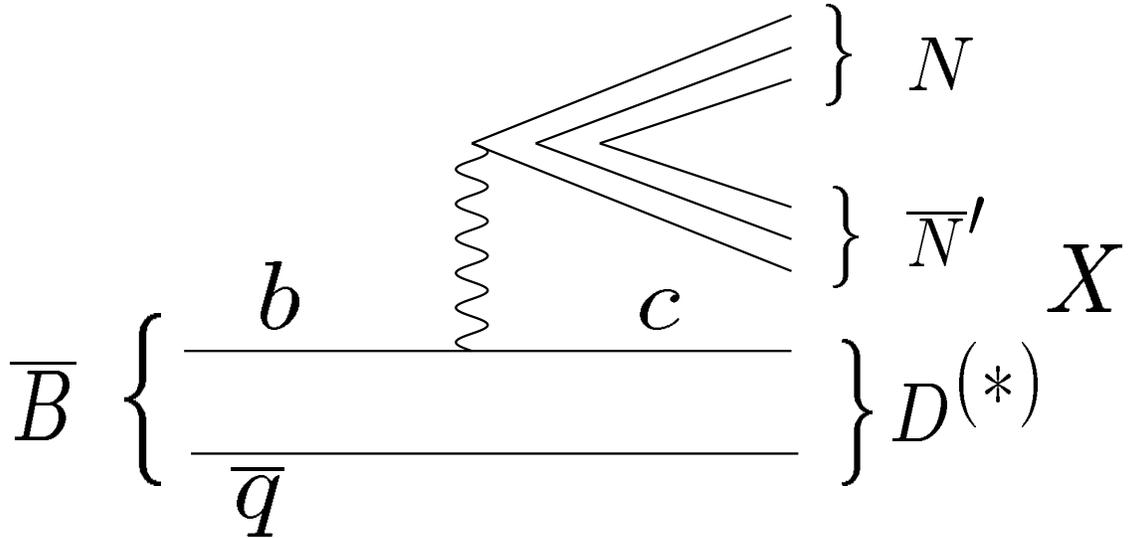}}}
\caption{Baryon production in $\protect\overline B$ decays, 
wherein the virtual $W$ hadronizes into a baryon- antibaryon pair.}

\end{figure}

\end{document}